\documentclass[preprint,3p,11pt]{elsarticle}

\usepackage{graphicx,amssymb}
\usepackage{epsf,psfig}

\journal{Astronomical \& Astrophysical Transactions}

\begin{document}

\begin{frontmatter}

\title{A rational galactic potential with accurate periodic orbits and quasi-integrals of motion}

\author{Euaggelos E. Zotos\fnref{}}

\address{Department of Physics, \\
Section of  Astrophysics, Astronomy and Mechanics, \\
Aristotle University of Thessaloniki \\
GR-541 24, Thessaloniki, Greece}

\fntext[]{Corresponding author: \\
E-mail: evzotos@physics.auth.gr}

\begin{abstract}
The motion in a simple, time independent rational galactic potential is studied. The potential is a generalization of a two dimensional harmonic oscillator potential and can be considered to describe plane motion in the central parts of a galaxy. There are cases, where the potential displays accurate periodic orbits, together with large chaotic regions. The starting position of these orbits can be found using analytical arguments. Using a quasi-integral of motion, we study the stability of the central periodic orbit and explain the degree of chaos in the inner regions of the galaxy. Our outcomes are compared with results coming from polynomial potentials with accurate periodic orbits and large chaotic regions.
\end{abstract}


\end{frontmatter}

\section{Introduction}

Galaxies are massive gravitationally bound systems, which consist of stars and stellar remnants, an interstellar medium of gas dust and a significant but poorly understood component, tentatively dubbed dark matter. Typical galaxies range from dwarfs with as few as ten million $(10^7)$ stars, up to giants with a hundred trillion $(10^{14})$ stars, all orbiting the galaxy's center of mass. Moreover, galaxies may contain star systems, star clusters and various interstellar clouds.

There are probably more than 170 billion galaxies in the observable universe. The majority of galaxies are organized into a hierarchy of associations called clusters, which in turn, can form larger groups called superclusters. These vast structures are generally arranged into sheets and filaments, which surround immense voids in the universe.

Although it is not yet well understood, dark matter appears to account for around $90\%$ of the total mass of most galaxies. Observational data suggest that, supermassive black holes may exist at the center of many, if not all, galaxies. They are proposed to be the primary cause of Active Galactic Nuclei (A.G.N), found at the core of some galaxies. Since the mass of gas and dust in most galaxies is less than $10\%$ of the total mass of the galaxy, it is a good approximation to consider galaxies as gravitationally bound assemblages of stars and dark matter.

Most galaxies in the universe are gravitationally bound to a number of other galaxies. These form a fractal-like hierarchy of clustered structures, with the smallest such associations being termed groups. A group of galaxies is the most common type of galactic cluster and these formations contain a majority of the galaxies (as well as most of the baryonic mass) in the universe. To remain gravitationally bound to such a group, each member of galaxy must have a sufficient low velocity, in order to prevent it from escaping (see Virial theorem). If there is sufficient kinetic energy, however, the group may evolve into a smaller number of galaxies through mergers.

Larger structures containing many thousands of galaxies packed into an area of a few megaparsecs across called clusters. Clusters of galaxies are often dominated by a single giant elliptical galaxy, known as the brightest cluster galaxy, which over time, tidally destroys its satellite galaxies and absorbers their mass to its own.

Superclusters contain tens of thousands of galaxies, which are often found in clusters, groups and sometimes individually. At the supercluster scale, galaxies are arranged into sheets and filaments surrounding vast empty voids. Above this scale, the universe appears to be isotropic and homogeneous.

In order to study the orbital motion of stars in a galaxy, we need a galactic dynamical model, which describes mathematically the motion of stars in the galaxy. A galactic dynamical model must give a general or global description of the galaxy. Furthermore, the outcomes from the dynamical model must agree with the data coming from observations. If this happens, one can say that the model describes satisfactorily the dynamical properties of the given galaxy. The reader can find interesting information regarding dynamical models of galaxies and potential-density pairs for spherical and flattened galaxies in Binney \& Tremaine (2008).

The aim of the present paper is: (i) to study numerically the motion in a rational galactic potential and to try to connect its nature with the dynamical parameters entering the system and (ii) to present some interesting theoretical arguments, which can be used, in order to support the numerical outcomes. The article is organized as follows: In Section 2, we present the dynamical model. The numerical outcomes of the research are given in Section 3. In Section 4, some useful theoretical results are presented, in order to explain and support the numerically found results. We close with a discussion and the conclusions of this research, which can be found in Section 5.

\section{Presentation of the dynamical model}

During the last decades, there has been a great deal of interesting research on chaotic behavior, integrability, resonances and related subjects for certain fairly simple analytic potentials. For perfectly valid reasons, much of this work (excellent summaries can be found in Lichtenberg and Lieberman, 1982 and in Tabor, 1984) is preoccupied with the use of the Poincar\'{e} surface of section (P.S.S) technique, the possible coexistence of additional isolating integrals or quasi-integrals of motion and their relation to the K.A.M theorem. In the author's view, this rather powerful mathematical machinery, tends to (unfortunately) overshadow some of the basic physical concepts involved and it is difficult to generalize results from one potential to related potentials.

It is well known that, if we expand a global potential near an equilibrium point, we get a potential of a two dimensional harmonic oscillator. This is the basic reason, why potentials made up of harmonic oscillators are the building blocks for local galactic type potentials. These potentials have been extensively used, for more than four decades (see Henon and Heiles, 1964; Saito and Ichimura, 1979; Deprit, 1991; Deprit and Elipe, 1991; Caranicolas, 1994; Ferrer et al., 1998a; Ferrer et al., 1998b; Elipe and Deprit, 1999; Elipe, 2000; Arribas et al., 2006) to model galactic motion. The general form of such a potentials is
\begin{equation}
V_A(x,y)=\frac{1}{2} \left[ \omega_1^2x^2 + \omega_2^2y^2 \right] + \epsilon V_1,
\end{equation}
where $V_1$ is a polynomial containing the perturbing terms, while $\epsilon$ is the perturbation strength.

In the present article we shall use the rational potential
\begin{equation}
V(x,y)=\frac{1}{2} \left[ \frac{x^2}{A+ \alpha_1 y^2} + \frac{y^2}{B + b_1 x^2} \right],
\end{equation}
where $A, B, \alpha_1, b_1$ are positive parameters. Potential (2) can be considered to describe plane motion in the central parts of a galaxy. There are several reasons for choosing potential (2). The first reason is that when $\alpha_1 = b_1 = 0$ potential (2) reduces to two dimensional harmonic oscillator potential, while for small values of $\alpha_1 \ll 1, b_1 \ll 1$, we can expand (2) in a Taylor series around the origin, keep the terms up to fourth order and find
\begin{equation}
V_T(x,y)=\frac{1}{2}\left[ \frac{x^2}{A} + \frac{y^2}{B} \right] - \frac{1}{2} \left[ \frac{\alpha_1}{A^2} + \frac{b_1}{B^2} \right] x^2 y^2 .
\end{equation}
Note that (3) is a potential of a two dimensional perturbed harmonic oscillator. In other words, potential (3) has the form (1) where
\begin{eqnarray}
\omega_1^2 &=& \frac{1}{A}, \nonumber \\
\omega_2^2 &=& \frac{1}{B}, \nonumber \\
\epsilon V_1 &=& -\frac{1}{2} \left[ \frac{\alpha_1}{A^2} + \frac{b_1}{B^2} \right] x^2 y^2.
\end{eqnarray}
The second reason is that, for larger values of $\alpha_1, b_1$ in equation, (2) we have a simple potential, which can be used in order to describe motion in a highly perturbed galaxy. A third reason is that, rational potentials made up of harmonic oscillators, if any, are not frequently used to describe motion in galaxies.

\section{Numerical outcomes}

The Hamiltonian corresponding to the potential (2) is
\begin{equation}
H = \frac{1}{2} \left( p_x^2 + p_y^2 \right) + \frac{1}{2} \left[ \frac{x^2}{A + \alpha_1 y^2} + \frac{y^2}{B + b_1 x^2} \right] = h,
\end{equation}
where $p_x, p_y$ are the momenta, per unit mass, conjugate to $x$ and $y$ and $h$ is the numerical value of the energy. Our numerical outcomes, are based on the numerical integrations of the equations of motion
\begin{eqnarray}
\frac{d^2x}{dt^2} &=& - \left[ \frac{1}{A + \alpha_1 y^2} + \frac{b_1 y^2}{\left(B + b_1x^2 \right)^2} \right] x, \nonumber \\
\frac{d^2y}{dt^2} &=& - \left[ \frac{1}{B + b_1 x^2}+ \frac{\alpha_1 x^2}{\left(A + \alpha_1 y^2 \right)^2} \right] y,
\end{eqnarray}
using a Bulirsch-St\"{o}er numerical integration Fortran code, with double precision subroutines. The accuracy of the calculations, were checked by the constancy of the energy integral (5), which was conserved up to the twelfth significant figure.

In order to visualize the regular or chaotic nature of motion, we obtain numerically the $x-p_x, y = 0, p_x > 0$ Poincar\'{e} phase plane, for the Hamiltonian (5). In order to keep things simple we shall consider the case, where we are in the 1:1 resonance, that is when $A = B$. Without loss of the generality, we take $A = B = 1$. The value of energy $h$ in all cases is equal to $1$. For small values of $\alpha_1, b_1$ the motion is regular and the chaotic regions, if any, are negligible.

For larger values of $\alpha_1, b_1$ considerable chaotic regions appear in the $(x,p_x)$ phase plane. The portrait of the phase plane when $\alpha_1 = 0.9, b_1 = 0.5$ is shown in Figure 1a. We can see a large chaotic sea, while the areas of regular motion are confined near the origin and around the stable periodic points on the $p_x$ axis. As we shall show in the next Section, these two elliptic periodic points represent accurate straight line periodic orbits going through the origin, when $\alpha_1 = b_1$. Figure 1b is same to Figure 1a, but when $\alpha_1 = 0.5, b_1 = 0.9$. Here, one observes again a large chaotic sea, while the regular regions seem to be smaller. The regular motion seems to be confined near the elliptic periodic points on the $p_x$ axis. What is more interesting, in this case, is that the central periodic point, which represents a periodic orbit, that is the $y$ axis, is now unstable, while in the case of Figure 1a this periodic orbit is stable. This orbital behavior will be explained in the next Section.
\begin{figure*}[!tH]
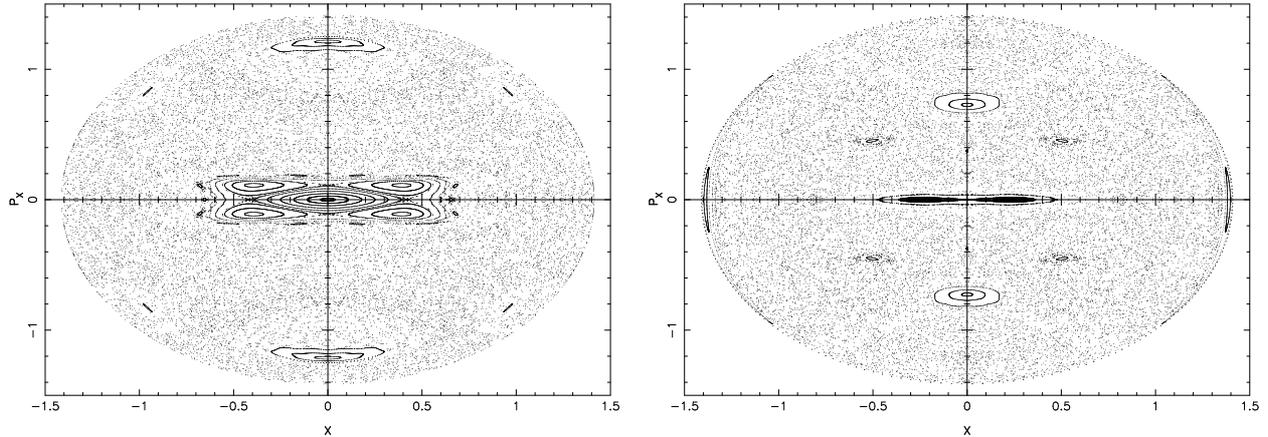

\centering
\resizebox{\hsize}{!}{\rotatebox{270}{\includegraphics*{Fig-1a.ps}}\hspace{1cm}
                      \rotatebox{270}{\includegraphics*{Fig-1b.ps}}}
\vskip 0.1cm
\caption{(a-b): The $(x,p_x)$ Poincar\'{e} phase plane when $A = B = 1, h = 1$, (a-left): $\alpha_1 = 0.9, b_1 = 0.5$. The $y$ axis is a stable periodic point and (b-right): $\alpha_1 = 0.5, b_1 = 0.9$. The $y$ axis is an unstable periodic point.}
\end{figure*}

Figure 2a which is a magnification of the central region of the $(x,p_x)$ phase plane of Fig. 1a, shows clearly that the central periodic point is stable. On the contrary, in Figure 2b which is a magnification of the central region of Fig. 1b, one can observe that in this case, the central periodic point is unstable. In fact, it produces a homoclinic chaotic region.
\begin{figure*}[!tH]
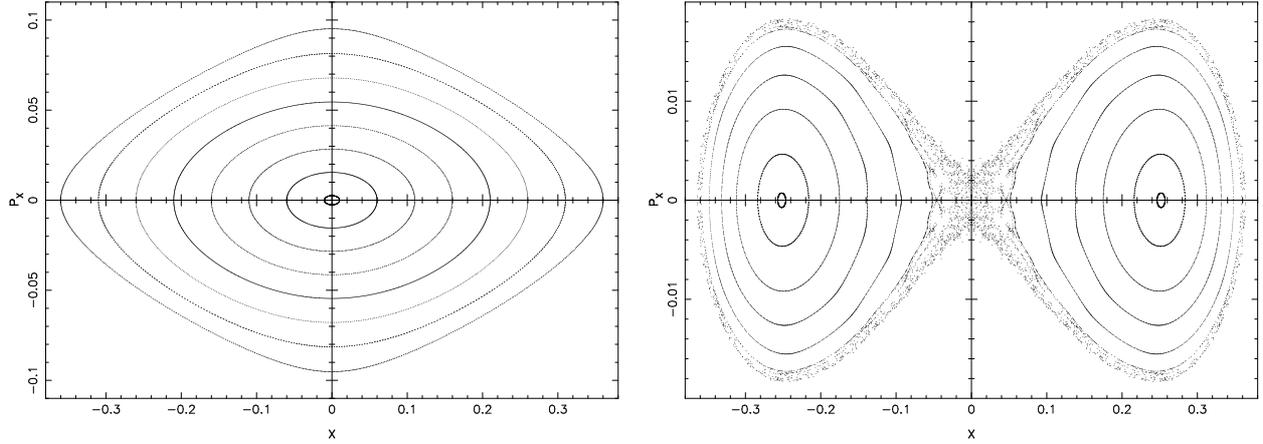

\centering
\resizebox{\hsize}{!}{\rotatebox{270}{\includegraphics*{Fig-2a.ps}}\hspace{1cm}
                      \rotatebox{270}{\includegraphics*{Fig-2b.ps}}}
\vskip 0.1cm
\caption{(a-b): Magnification of the central region of the $(x,p_x)$ phase planes of Figs. 1a and 1b respectively.}
\end{figure*}

\section{Theoretical approach}

In what follows, we shall present some theoretical calculations in order to support our numerical results of the previous Section. First we shall show, that there are cases where the dynamical system (2) displays accurate periodic orbits.

We write the equations of motion (6) in the form
\begin{eqnarray}
\dot{x} &=& p_x, \nonumber \\
\dot{y} &=& p_y,
\end{eqnarray}
\begin{eqnarray}
\dot{p_x} &=& - \left[ \frac{1}{A + \alpha_1 y^2} - \frac{b_1 y^2}{\left( B + b_1x^2 \right)^2} \right] x, \nonumber \\
\dot{p_y} &=& - \left[ \frac{1}{B + b_1 x^2} - \frac{\alpha_1 x^2}{\left( A + \alpha_1 y^2 \right)^2} \right]y,
\end{eqnarray}
where the quantities in the brackets can be considered as the squares of the frequencies of oscillation along the $x$ and $y$ axes respectively and the dot indicates derivative with respect to time. One observes that when $A = B$ and $\alpha_1 = b_1$, for the motion along the straight lines
\begin{equation}
x = \pm y \ \ \ ,
\end{equation}
the two frequencies of oscillations become equal, while the pairs of equations (7) and (8) are identical. On this basis, (9) is a solution of the system of differential equations (7)-(8) giving the 1:1 resonant straight line periodic orbits going through the origin. Note that this happens only, when $A = B$ and $\alpha_1 = b_1$. Furthermore, we must emphasize that if we use the Taylor expansion (3) for the potential (2), the accurate periodic orbits (9) are present for all values of $\alpha_1$ and $b_1$ (see Caranicolas and Innanen, 1992). This happens because information is lost through the expansion of the rational potential (2) in a Taylor series.

Inserting the values of (7) and (9) in the Hamiltonian (5) and after setting $A = B$ and $\alpha_1 = b_1$ we obtain
\begin{equation}
H = p_x^2 + \frac{x^2}{A + \alpha_1 x^2} = h.
\end{equation}
Using the origin, as a starting point, we find the initial conditions of the straight line periodic orbits, from (10), which gives
\begin{equation}
x = y = 0, \ \ p_x = p_y = \sqrt{h}.
\end{equation}

Let us now come to explain the behavior of orbits in the central region of the $(x,p_x)$ phase plane. This behavior can be explained using a quasi-integral of motion, which is a generalization of the particle´s energy in the direction of the $x$ axis (see Binney \& Tremaine, 2008). As such a quasi-integral we take
\begin{equation}
f(x,p_x) = \frac{1}{2} \left(x^2 + 10p_x^2 \right)
+ 5(\alpha_1 - b_1) \left[\lambda x^2 + p_x^2x^2 + k p_x^2 \right],
\end{equation}
where $k, \lambda$ are positive parameters. Relation (12) was found, using a combination of theoretical analysis along with numerical simulations of the dynamical system and describes the structure and topology of the $(x,p_x)$ Poincar\'{e} phase plane, only near the central region. In order to keep things simple, we define a range of values for the parameters, which is $\alpha_1 \leq 0.9, b_1 \leq 0.9, \alpha_1 + b_1 \leq 1.4$. Figure 3a shows the curves
\begin{equation}
f(x,p_x) = c,
\end{equation}
where $c = $ 0.0015, 0.012, 0.035, 0.07, 0.115, 0.17, 0.24. The values of the other parameters are: $\alpha_1 = 0.9, b_1 = 0.5, k = 0.1, \lambda = 0.5$. We observe that the origin is a stable periodic point. Figure 3b is same to Fig. 3a but when $\alpha_1 = 0.5, b_1 = 0.9$ and $c = $ -0.00001, -0.002, -0.007, -0.0155, -0.0275, -0.043, -0.062. Now we see that the origin has become an unstable periodic point. Note that these results, agree with the numerically obtained results, shown in Figs. 2a and 2b.
\begin{figure*}[!tH]
\centering
\resizebox{0.80\hsize}{!}{\rotatebox{0}{\includegraphics*{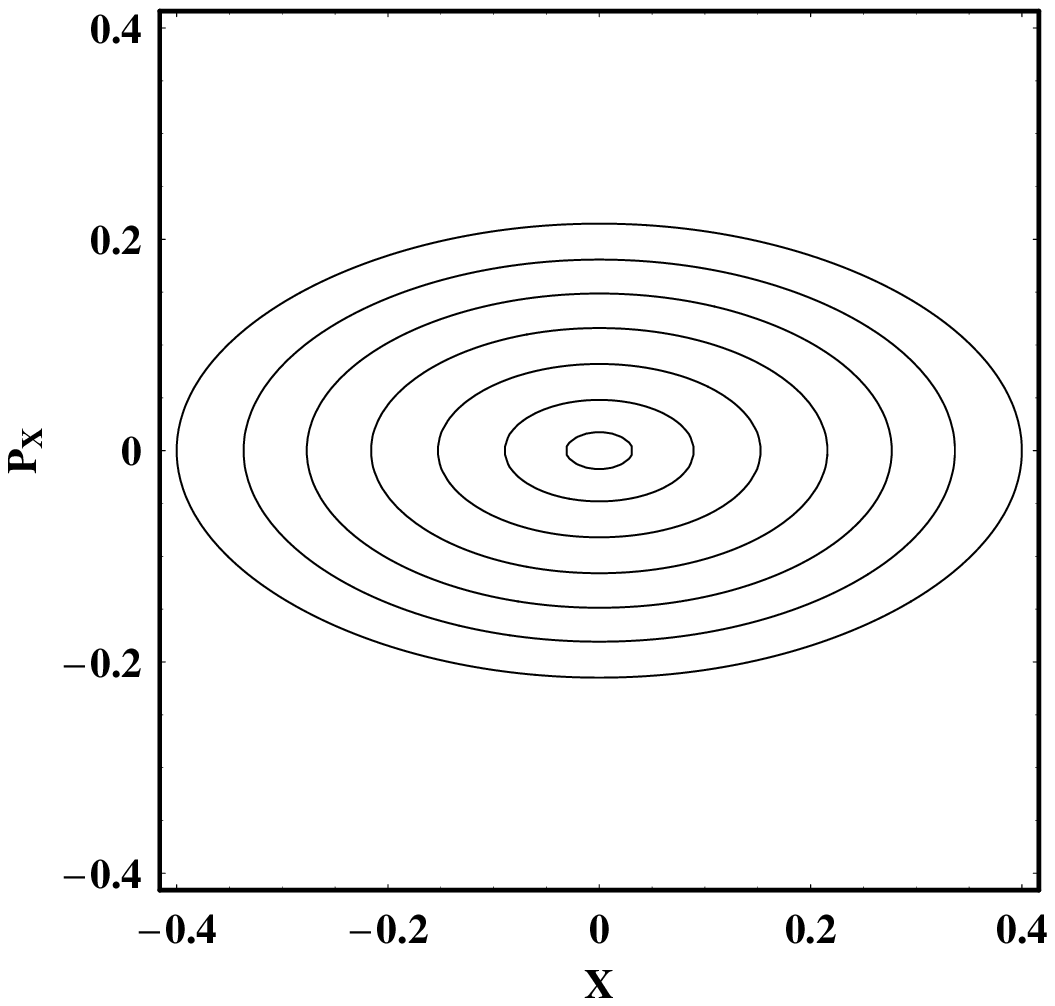}}\hspace{1cm}
                          \rotatebox{0}{\includegraphics*{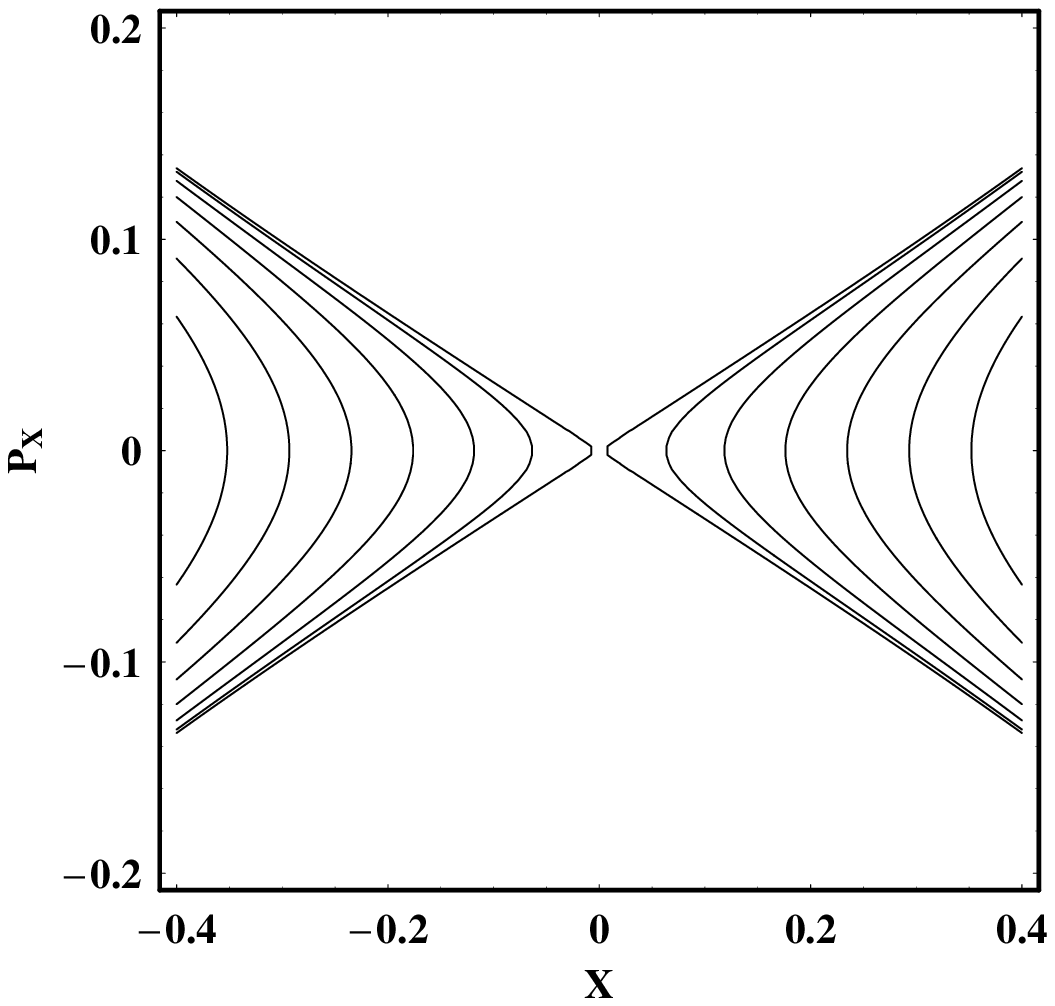}}}
\vskip 0.1cm
\caption{(a-b): Topology and structure of the region near the center of the phase plane, obtained using the quasi-integral (12). The values of the parameters are as in Fig. 1a (a-left) and Fig. 1b (b-right).}
\end{figure*}

The exchange of stability of the origin, strongly depends on the particular values of $\alpha_1$ and $b_1$. In what follows, we shall show this dependence analytically using (12). It is well known, that the maxima and minima of (12) indicate stable periodic points, while saddle points indicate unstable periodic points. Thus we have
\begin{eqnarray}
\frac{\partial f}{\partial x} &=& x + 10 \left( \alpha_1 - b_1 \right) x \left(p_x^2 + \lambda \right) = 0, \nonumber \\
\frac{\partial f}{\partial p_x} &=& 10p_x \left[ 1 + 10 \left( \alpha_1 - b_1 \right) \left(x^2 + k \right) \right] = 0.
\end{eqnarray}

We are interesting in the solution
\begin{equation}
x_0 = p_{x0} = 0,
\end{equation}
of the system of algebraic equations (14), while we ignore all the other solutions, due to the fact that they correspond to complex numbers for the particular values of the parameters $(\alpha_1, b_1, k, \lambda)$. Solution (15) represents the central invariant point, that is the $y$ axis. We also find from (14) the following
\begin{eqnarray}
A &=& \frac{\partial^2f}{\partial x^2} = 1 + 10 \left(\alpha_1 - b_1 \right) \left(p_x^2 + \lambda \right), \nonumber \\
B &=& \frac{\partial^2f}{\partial p_x^2} = 10 \left[1 + \left(\alpha_1 - b_1 \right) \left(x^2 + k \right) \right], \nonumber \\
\Gamma &=& \frac{\partial^2f}{\partial x \partial p_x} = 20 \left(\alpha_1 - b_1 \right) x p_x.
\end{eqnarray}

Setting solution (15) in equations (16), we find
\begin{eqnarray}
A_0 &=& A\left(x_0,p_{x0}\right) = 1 + 10 \lambda \left(\alpha_1 - b_1 \right), \nonumber \\
B_0 &=& B\left(x_0,p_{x0}\right) = 10 \left[1 + k \left(\alpha_1 - b_1 \right) \right], \nonumber \\
\Gamma_0 &=& \Gamma \left(x_0,p_{x0}\right) = 0.
\end{eqnarray}

The $y$-axis is a stable periodic if
\begin{equation}
\Delta = A_0 B_0 - \Gamma_0 ^2 > 0.
\end{equation}
From equations (17) we have
\begin{equation}
\Delta = 10 \left[1 + 10 \lambda \left(\alpha_1 - b_1 \right) \right] \left[1 + k \left(\alpha_1 - b_1 \right) \right].
\end{equation}
It is evident that when $b_1 \leq \alpha_1$ the $y$-axis is a stable periodic orbit.

The $y$ axis is an unstable periodic orbit, when the quantity $\Delta$ is negative. In this case, for the adopted values of $k, \lambda$ and a given value of the parameter $b_1$, we solve the corresponding inequality. For instance, when: $k = 0.1, \lambda = 0.5, b_1 = 0.9$ the corresponding inequality is
\begin{equation}
\Delta = 5 \alpha_1^2 + 42 \alpha_1 -31.85 < 0,
\end{equation}
which holds when $-9.1 < \alpha_1 < 0.7$. Note that, because we have assumed only positive values of $\alpha_1$, we take $0 < \alpha_1 < 0.7$.

\section{Discussion and conclusions.}

In this work, we have used a simple rational potential, in order to describe plane motion in the central parts of a galaxy. This rational potential can be considered as a generalization of the potentials made up of harmonic oscillators, which have been used for more than four decades to model galactic motion. What is new and interesting here, is that the non-linear terms are introduced in a different way. Therefore, we present a different path of inserting the perturbing terms in a galactic potential. In the present article, we have decided to study the case when $A = B = 1$, that is, the case where the system displays the 1:1 resonance.

The outcomes of this research come out of two sources. The first source is the numerical integration, while the second source is some elementary analytical calculations. The numerical calculations show that when $\alpha_1 \ll 1, b_1 \ll 1$ the motion is regular, while for larger values of the above parameters the systems shows large chaotic regions on the $(x,p_x)$ phase plane. The regular motion is confined mainly, near the stable periodic points on the $p_x$ axis. A small regular region appears near the $x=p_x=0$ periodic point, which represents a periodic orbit which is the $y$ axis. This small regular region appears only when $\alpha_1 \geq b_1$.

Of particular interest is the case, when $\alpha_1 = b_1$, where the system displays accurate periodic orbits. These periodic orbits are straight lines going through the origin. The existence of those periodic orbits is shown analytically, using the equations of motion. We emphasize that, in this case, we have a dynamical model, which produces accurate periodic orbits together with large chaotic regions. This phenomenon is similar to the case studied by Caranicolas (2000), where accurate periodic orbits were constructed in polynomial potentials using the inverse problem theory. Those potentials displayed also large chaotic regions. Here, we see that the co-existence of chaos and accurate periodic orbits, appears not only in polynomial, but also in rational potentials.

The stability of the $y$ axis can be studied using a quasi-integral of motion, which can be considered as a generalization of the energy along the $x$ axis. This quasi-integral was constructed using a combination of theoretical and numerical analysis. Note that the quantity (12), is a local quasi-integral and as such a local quasi-integral is used to study local dynamical properties, that is only around central region of the origin $x = p_x = 0$ of the phase plane. Comparison of theoretical and numerical outcomes show that this quasi-integral explains satisfactorily the behavior of orbits near the origin.

Before closing, the author would like to make a comment. Strictly speaking, the results of Section 4 are not pure analytical, but rather semi-analytical. It is true, that semi-analytical methods are frequently used in Celestial Mechanics (see Henrard, 1990) and Galactic Dynamics (see Caranicolas and Innanen, 1992; Caranicolas, 1993), as a very sharp tool, during the last decades.

\end{document}